\pdfoutput=1

\documentclass[
 reprint,
 nofootinbib,
 amsmath,amssymb,
 aps,
 prb,
]{revtex4-1}

\usepackage{graphicx}
\usepackage{xcolor}
\usepackage{transparent}
\usepackage{dcolumn}
\usepackage{bm}	
\usepackage{amssymb}
\usepackage{braket}
\usepackage{mathtools}
\usepackage{setspace}
\usepackage{amsmath}
\usepackage{esdiff}
\usepackage[sort&compress]{natbib}
\usepackage{caption}
\usepackage{subcaption}
\usepackage{fancyhdr}
\usepackage{titlesec}

\begin{document}

\title{Method of images applied to driven solid-state emitters}

\author{Dale Scerri}
\email{ds32@hw.ac.uk}
\author{Ted S. Santana}
\author{Brian D. Gerardot}
\author{Erik M. Gauger}
\email{e.gauger@hw.ac.uk}
 
\affiliation{SUPA, Institute of Photonics and Quantum Sciences, Heriot-Watt University, EH14 4AS, United Kingdom.}
\date{\today}

\begin{abstract}
Increasing the collection efficiency from solid-state emitters is an important step towards achieving robust single photon sources, as well as optically connecting different nodes of quantum hardware. A metallic substrate may be the most basic method of improving the collection of photons from quantum dots, with predicted collection efficiency increases of up to 50\%. 
The established `method-of-images' approach models the effects of a reflective surface for atomic and molecular emitters by replacing the metal surface with a second fictitious emitter which ensures appropriate electromagnetic boundary conditions. Here, we extend the approach to the case of driven solid-state emitters, where exciton-phonon interactions play a key role in determining the optical properties of the system. We derive  an intuitive polaron master equation and demonstrate its agreement with the complementary half-sided cavity formulation of the same problem. Our extended image approach offers a straightforward route towards studying the dynamics of multiple solid-state emitters near a metallic surface.
\end{abstract}

\maketitle

\section{\label{intro}Introduction}

The problem of a dipole emitter placed close to a reflective surface has received much interest over the last few decades: seminal work  \cite{DREXHAGE} by Drexhage in 1970 first demonstrated that a reflective interface modifies the intrinsic properties of the emitter, influencing both the emission frequency \cite{frequencyshift,Morawitz1969} and the emitter's excited lifetime \cite{Morawitz1969, babiker, babiker2, babiker:superradiance, ficek2005quantum}. Recently, a sound analogue of Drexhage's experiment has been performed to study the acoustic frequency shifts of a gong struck near a hard wall \cite{acoustic}.

Mirrors have widespread use for directing light from sources that emit across a extended solid angle, for example in the form parabolic reflectors in everyday light sources. On the nanoscale, precise guiding of photons into particular optical modes is of paramount importance for quantum information processing and communication, where on demand single photons are required \cite{KLM, SinglePhotonTransport1, SinglePhotonTransport2, SinglePhotonTransport3}. 
Although micron-sized spherical mirrors for open access microcavities \cite{JasonSmith} have recently enabled the investigation of quantum dot--cavity systems in the strong coupling regime \cite{Warburton,Imamoglu}, the use of sophisticated mirrors remains a challenge for solid-state quantum emitters that are often embedded in heterogenous layers of substrates with varying refractive indices. This motivates the more straightforward alternative of increasing the photon collection efficiency by placing the emitter above a planar mirroring interface \cite{SurfaceEnhancedRef1, SurfaceEnhancedRef2, SurfaceEnhancedRef3}. Interestingly, the presence of even such a simple mirror also affects the physical properties of the emitter, as discussed above.

In recent years, progress in the synthesis and control of solid-state emitters has enabled experimental investigation of these modified properties of condensed-state emitters including quantum dots (QDs) \cite{nanowire, gerardot} as well as perovskite \cite{imageexcitons} and transition metal dichalcogenide monolayers \cite{MonolayerRef} deposited on reflective surfaces. Circuit QED analogues of an atom and a variable mirror have also been successfully implemented \cite{circuit_qed, circuit_qed2}; these offer the advantage of increased control over the artificial atom's interaction with the mirror. With improved atom-mirror coupling, Hoi \textit{et al.} managed to collect over $99\%$ of the radiation by coupling a transmon microwave emitter to a 1D superconducting waveguide \cite{circuit_qed}. 

\begin{figure}[t!]
\begin{center}
\includegraphics[width=\linewidth]{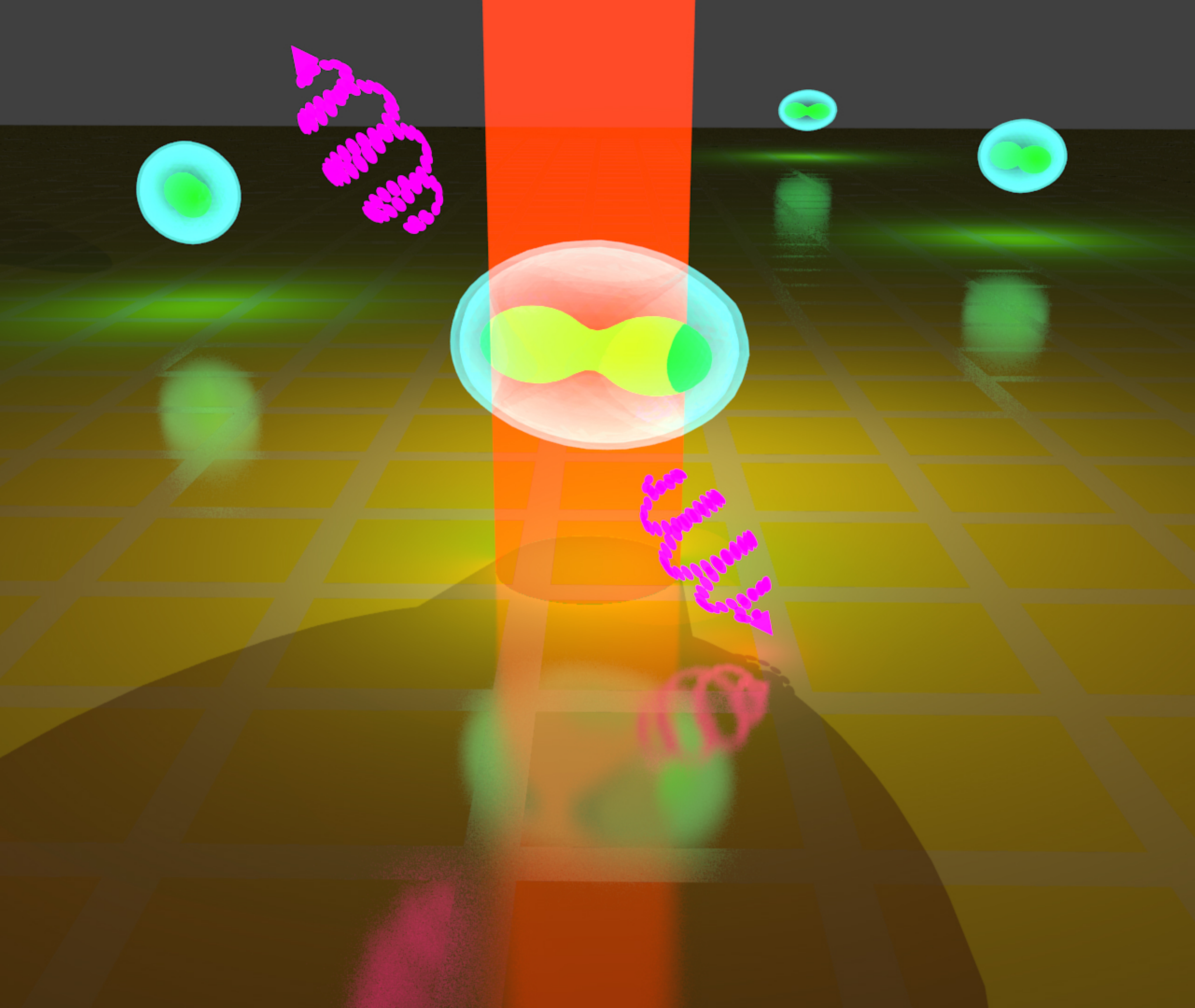}      
\caption{Artistic rendition of a driven quantum dot (QD), depicted as a cyan spheroid, in the proximity of a golden metallic surface. The corresponding `image dot' is shown blurred on the other side `below' of  the semiconductor-gold interface. The optical dipoles are depicted as `dumbbells' within the QDs. The vertical red beam represents the laser driving, and the magenta spiralling arrows indicate scattered photons.}
\label{blender}
\end{center}
\end{figure}

\begin{figure}[t!]
\centering
\def\svgwidth{0.4\textwidth}
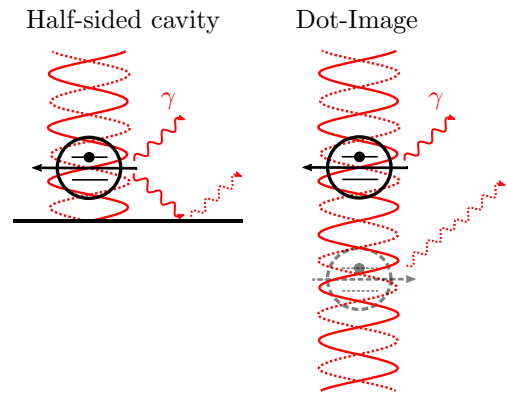 
\caption{Two equivalent descriptions of an emitter near a perfect metallic mirror. {\bf Left:} schematic of the Green's function and half-sided cavity approaches. {\bf Right:} the emitter supplemented with a fictitious image dipole. The solid (dashed) red arrows indicate emitted (reflected) photons whereas the solid (dashed) red curve indicates the incident (reflected) driving beam.}
\label{schematic}
\end{figure}

Several theoretical investigations \cite{frequencyshift, Morawitz1969, babiker, ficek2005quantum} have shown that an atomic two-level system (TLS) near a reflective surface can be  modelled as a pair of emitters: the real one as well as an identical emitter that is placed equidistant from, but on the opposite side of, the interface (see Figs.~\ref{blender} and \ref{schematic}). The basic idea follows that of the electrostatics concept of an image charge to capture the surface charge distribution that ensures meeting the electric field boundary conditions \cite{jackson}. In the optical case, the `method of images' relies on considering the emission from the combined dipole-image system. This yields the same expression for the modified spontaneous emission (SE) rate which one obtains from a full QED treatment (employing surface-dependent response functions to arrive at the modifications to the emitter's lifetime and transition frequency)\cite{agarwal3}.  The image dipole treatment has also been applied to model the surface-induced modifications of more complex structures such as molecules \cite{sers, barnesreview}, multiple dipole emitters \cite{George1985,imagedipoleold2,sanders} and solid state-emitters \cite{nanowire,imageexcitons}. To date, however, the latter have largely ignored the vibrational solid state environment and the continuous wave (cw) laser driving typical of a resonance fluorescence (RF) setting.

Motivated by these successes, we here present a full image dipole polaron master equation (ME) treatment of a driven TLS (such as, e.g., a quantum dot) in the proximity of a metal surface (see Fig.~\ref{blender}). Our calculations extend previous image dipole studies as follows: (i) we consider driven systems, showing how to incorporate a laser driving term into the dipole and image Hamiltonian; (ii) we discuss the need for introducing an additional `selection rule' to prevent unphysical double excitation; (iii) we demonstrate how a solid-state phonon environment can be accounted for -- via a single bosonic bath that is perfectly correlated across the real emitter and its image. 

We will show that the resulting master equation model remains highly intuitive and possesses appealing simplicity. We establish the correctness of this model by comparing its results to those obtained from an alternative calculation which does not involve fictitious entities or rely on ad-hoc assumptions: the half-sided cavity model. This agreement gives us confidence that the model could also be extended to the case of multiple solid-state emitters near a reflective surface, laying the groundwork for the investigation of collective effects in this setting, where we believe that an image approach will be easier to deploy than both the Green's function and the half-sided cavity approach.

This Article is organised as follows: We will start by briefly summarising the results from the established Green's function method for calculating the SE rate of a `bare' dipole emitter. Next, we shall derive a ME for the emitter by treating the metal surface as a half-sided Fabry--P{\'e}rot cavity, providing the benchmark model for a single TLS near the metal surface (see Fig.~\ref{schematic}a). Finally, we formulate the ME using the method of images (see Fig.~\ref{schematic}b). We show that, with suitable alterations, the two-body ME reduces to an effective two level system with rates and energy shifts agreeing with the cavity model. 
Finally, we put our model to use to obtain the RF spectrum of the modified system, featuring a phonon sideband, the Mollow triplet, and the ratio of coherently to incoherently scattered light. 

\section{\label{Green}Green's function approach: Brief summary}

We begin by summarising the main results of the Green's function approach for modelling the optical environment of a dipole emitter. This can be applied to obtain the SE rate of an emitter in free space \cite{principles} as well as in the presence of a metallic surface \cite{principles,green,babiker2}. Whilst this approach gives a closed analytical solution for the case of a single dipole, a numerical route has to be taken to model a system comprised of a larger number of emitters \cite{sanders, principles}, even in the absence of a driving field and phonon-environments. Therefore, we here limit the discussion to a single `bare' emitter as an independent reference point for the SE rate (and energy shift) in that idealised configuration.

Let the dipole be situated at position $\mathbf{r}_d$, where $\mathbf{r}_d$ is perpendicular to a metal surface containing the origin of the coordinate system. In the Green's function approach, the emitter is usually modelled as a classical dipole oscillating harmonically with amplitude $\mathbf{x}$ at frequency $\omega_0$ about $\mathbf{r}_d$ \cite{sanders}. In vacuum, the SE rate can be calculated as
\begin{equation}\label{SEGreen}
\gamma^{pt}_0(\omega_0) = \frac{4 \omega^2_0}{ \pi \epsilon_0 \hbar c^2}\left[ \hat{\mathbf{d}} \cdot \mathrm{Im}\{ \mathbf{G}(\mathbf{r}_d, \mathbf{r}_d; \omega_0) \} \cdot \hat{\mathbf{d}} \right]~,
\end{equation}
where $\epsilon_0$ is the electric permittivity of vacuum, $c$ is the speed of light, $\hat{\mathbf{d}}$ is a unit vector indicating the direction of the emitter's dipole moment, and $\mathbf{G}(\mathbf{r}_d, \mathbf{r}_d; \omega_0)$ is the Fourier transform of the dyadic Green's function at the emitter's position \cite{principles}. In Ref.~\cite{sanders}, Choquette \textit{et al.} studied the the collective decay rate of $N$ such classical emitters near a planar interface, arriving at a diagonal Green's function matrix, so that Eq.~\eqref{SEGreen} allows one to find the SE rate for arbitrary dipole orientations. 

To obtain the SE rate in a dielectric environment, we consider the following expression for the normalised dissipated power:
\begin{equation}\label{power}
\frac{P}{P_0} = 1 + \frac{6 \pi \epsilon_0 \epsilon_r}{|\mathbf{d}|^2 k^3} \mathrm{Im} \{ \mathbf{d}^* \cdot \mathbf{E}_s(\mathbf{r}_d) \} ~,
\end{equation}
where $P_0$ is rate of energy dissipation in free space, $\epsilon_r$ and $k$ are the relative permittivity and wave vector magnitude in the dielectric surrounding the emitter, respectively, and $\mathbf{E}_s(\mathbf{r}_d)$ is the scattered electric field at the dipole's position (which, for a single dipole near the surface, corresponds to the reflected field) \cite{principles}. The connection between the Green's function and the decay rate of the dipole emitter is established via the relationship
\begin{equation}
\frac{P}{P_0} = \frac{\gamma^{pt}(\omega_0)}{\gamma^{pt}_0(\omega_0)} ~.
\end{equation}
Rearranging the above then yields an integral expression for the desired SE rate $\gamma^{pt}(\omega_0)$.

We note that the Green's function method is not limited to ideal metallic interfaces but can also be applied straightforwardly to reflective dielectric interfaces, simply by substituting appropriate dielectric constants into the above relevant expressions \cite{principles}. In this case, one obtains qualitatively very similar results for a dielectric mirror, especially at larger separations \cite{principles}. Whilst the method of images fundamentally relies on the assumption of a perfectly conducting interface, it is fair to assume its qualitative predictions will by analogy also carry across to the case of dielectric mirrors.

\section{\label{halfcavity}Half-sided Cavity Model}

\begin{figure}[t!]
\centering
\def\svgwidth{0.7\columnwidth}
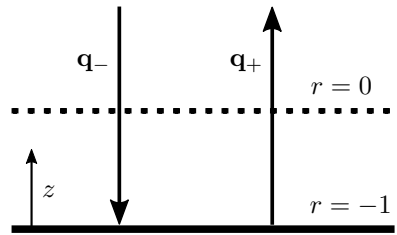
\caption{The limiting case of the Fabry--P{\'e}rot cavity, effectively reducing to a single perfectly reflecting surface. The arrows indicate the wavevectors in \eqref{cavityRabi} and \eqref{cavity_spatial_fns}, and $r$ denotes the surface reflection coefficient \cite{demartini, ficek2005quantum}.}
\label{cavityvecs}
\end{figure}

In the previous section, we discussed how to determine the SE rate for an undriven emitter interacting only with a photonic environment. However, in order to fully model a solid-state emitter such as a QD, we need to include interactions between the emitter and its phonon environment \cite{Machnikowski:Phonons,Machnikowski:Rabi}. Now we shall derive the polaron ME for a TLS near a metal surface, by modelling the latter as a half-sided Fabry--P{\'e}rot cavity positioned at $z=0$ lying in the $xy$ plane, and the QD positioned at $z = r_d \geq 0$, where $r_d = |\mathbf{r}_d|$. Our calculation follows the general cavity model from Refs.~\cite{demartini, ficek2005quantum}, taking the appropriate limits for the reflectivity and transmittivity of the two mirrors to obtain, effectively, only a single perfectly reflecting surface (see Fig.~\ref{cavityvecs}).

\subsection{\label{halfcavity:hamiltonian}Hamiltonian}

We consider a driven TLS with ground state $\Ket{0}$ and excited state $\Ket{X}$, which is governed by the following Hamiltonian in a rotating frame and after the usual rotating wave approximation ($\hbar = 1$)

\begin{equation}\label{CavityHamiltonian}
H_S = \delta \Ket{X} \Bra{X} + \frac{\Omega^*_{cav}}{2}\Ket{0} \Bra{X} + \mathrm{H.c.} ~,
\end{equation}
where H.c.~denotes the Hermitian conjugate and  $\delta = \omega_0 - \omega_l$ is the detuning between the TLS transition frequency $\omega_0$ and the laser frequency $\omega_l$. $\Omega_{cav}$ is the effective Rabi frequency in the presence of the metal surface, given by
\begin{equation}\label{cavityRabi}
\Omega_{cav} = 2\sqrt{\frac{\omega_l}{2 \epsilon V}}~ \mathbf{d} \cdot \left( \mathbf{e}_{l_-} \mathrm{e}^{-i \mathbf{q}_{l} r} - \mathbf{e}_{l_+} \mathrm{e}^{i \mathbf{q}_{l} r} \right) ~,
\end{equation}
where $\mathbf{q}_l$ is the laser field wavevector, with polarisation $\mathbf{e}_{l_-}$ ($\mathbf{e}_{l_+}$ after reflection), as shown in Fig.~\ref{cavityvecs} for the case of the laser  beam being perpendicular to the surface. Photon and phonon environments are modelled by the Hamiltonians
\begin{align}
H^{pt}_E &= \sum_{\mathbf{q}, \, \lambda} \nu_\mathbf{q} a^\dagger_{\mathbf{q}\lambda} a_{\mathbf{q}\lambda}~, \\
H^{pn}_E &= \sum_{\mathbf{k}} \omega_\mathbf{k} b^\dagger_\mathbf{k} b_\mathbf{k} ~,
\end{align}
where $b^\dagger_\mathbf{k}$ and $a^\dagger_{\mathbf{q}\lambda}$ ($b_\mathbf{k}$ and $a_{\mathbf{q}\lambda}$) are the $\mathbf{k}$-phonon and $\mathbf{q}\lambda$-photon creation (annihilation) operators, respectively. In the dipole approximation, the photon interaction Hamiltonian is of the form
\begin{equation}
H^{pt}_I =  -\mathbf{d} \cdot \mathbf{E}(\mathbf{r}_d) (\Ket{0} \Bra{X} + \Ket{X} \Bra{0}) ~
\label{eq:hpt0}
\end{equation}
with $\mathbf{E}(\mathbf{r})$ being the Schr{\"o}dinger picture electric field for the half-sided cavity \cite{ficek2005quantum, demartini}, 
\begin{equation}\label{electricfield}
\mathbf{E}(\mathbf{r}) = i \sum_{\mathbf{q}, \lambda} \left[ \mathbf{u}_{\mathbf{q} \lambda}(\mathbf{r}) a_{\mathbf{q} \lambda} - \mathrm{H.c.} \right] ~.
\end{equation}
The spatial mode functions $\mathbf{u}_{\mathbf{q} \lambda}(\mathbf{r})$ for an ideal half-sided cavity (of perfect reflectivity) are given by

\begin{equation}\label{cavity_spatial_fns}
\mathbf{u}_{\mathbf{q} \lambda}(\mathbf{r}) = \sqrt{\frac{\omega_{\mathbf{q} \lambda}}{2 \epsilon V}}\left( \mathbf{e}_{\mathbf{q}_- \lambda} \mathrm{e}^{i \mathbf{q}_- r} - \mathbf{e}_{\mathbf{q}_+ \lambda} \mathrm{e}^{i \mathbf{q}_+ r} \right)~.
\end{equation}
Here, $\mathbf{q}_-$ ($\mathbf{q}_+$) is the incident (reflected) wavevector, with corresponding polarisation $\mathbf{e}_{\mathbf{q}_- \lambda}$ ($\mathbf{e}_{\mathbf{q}_+ \lambda}$). For simplicity, we have assumed that the dipole moment $\mathbf{d}$ of the TLS is real. 

The interaction with the phonon bath can be generically represented by the Hamiltonian \cite{Mahan}
\begin{equation}
H^{pn}_I = \Ket{X} \Bra{X}\sum_{\mathbf{k}} g_\mathbf{k} ( b^\dagger_\mathbf{k} + b_\mathbf{k} ) ~,
\end{equation}
where $g_\mathbf{k}$ is the coupling strength of the TLS's excited electronic configuration with phonon mode $\mathbf{k}$. We move to the polaron frame by employing the standard Lang--Firsov-type transformation $U = e^S$,  $S = \Ket{X}\Bra{X} \sum_{\mathbf{k}} (g_\mathbf{k} / \omega_\mathbf{k}) ( b^\dagger_\mathbf{k} - b_\mathbf{k} )$, obtaining the following transformed system Hamiltonian:
\begin{align}\label{polaronsystem}
\begin{split}
H_{SP} = \delta' \Ket{X} \Bra{X} &+ \frac{\Omega^*_{cav}}{2}\Ket{0} \Bra{X} B_-  \\
						 &+ \frac{\Omega_{cav}}{2}\Ket{X} \Bra{0} B_+~,
\end{split}
\end{align}
where $\delta' = \delta - \sum_\mathbf{k} g^2_\mathbf{k} / \omega_\mathbf{k}$ (becoming $\delta - \int_0^\infty J_{pn}(\omega) / \omega$ in the continuum limit), and the phonon bath operators $B_\pm$ are defined as $B_\pm = \Pi_\mathbf{k} D_\mathbf{k} (g_\mathbf{k} / \omega_\mathbf{k})$, with $D_\mathbf{k}(\pm \alpha) = \exp[\pm(\alpha b^\dagger_\mathbf{k} -\alpha^* b_\mathbf{k})]$ being the $\mathbf{k}$th mode displacement operator. For numerical results we shall later use a superohmic exciton-phonon spectral density $J_{pn}(\omega)$ with exponential cut-off at frequency $\omega_c$ that is appropriate for self-assembled III-V quantum dots \cite{Ramsay, phononrabi2}: 
\begin{equation}\label{phonon_spectral_density}
J_{pn}(\omega) = \alpha \omega^3 \mathrm{e}^{-\frac{\omega^2}{\omega_c^2}}~.
\end{equation}

In the polaron frame the light-mattter interaction Hamiltonian Eq.~\eqref{eq:hpt0} becomes 
\begin{align}
\begin{split}
H^{pt}_{IP} = &i \Ket{0} \Bra{X} B_-  \sum_{\mathbf{q}, \lambda} \mathbf{d}\cdot\mathbf{u}^*_{\mathbf{q}\lambda}(\mathbf{r}_d) a^\dagger_{\mathbf{q}\lambda} \\
                    -&i \Ket{X} \Bra{0} B_+  \sum_{\mathbf{q}, \lambda} \mathbf{d}\cdot\mathbf{u}_{\mathbf{q}\lambda}(\mathbf{r}_d) a_{\mathbf{q}\lambda} ~.
\end{split}
\end{align}
With the definitions $A_1^{pt} = \Ket{0} \Bra{X}$, $A_2^{pt} = A_1^{pt \dagger}$, $B^{pt}_{1/2} \equiv B_\mp$, $C_1 = i \sum_{\mathbf{q}, \lambda} \mathbf{d}\cdot\mathbf{u}^*_{\mathbf{q}\lambda}(\mathbf{r}_d) a^\dagger_{\mathbf{q}\lambda}$, and $C_2 = C^\dagger_1$, we can compactly write the above Hamiltonian as
\begin{equation}
\label{eq:compactHpti}
H^{pt}_{IP} = \sum_{i=1}^2 A^{pt}_i \otimes B^{pt}_i \otimes C_i ~,
\end{equation}
Since the second term in Eq.~\eqref{polaronsystem} contains system and environment operators, we identify this as our new exciton-phonon interaction term \cite{phononreview}. This new interaction term possesses a non-zero expectation value with respect to the thermal equilibrium bath state $\rho^{pn}_E$; tracing out the phonon bath degrees of freedom, we thus obtain
 \begin{align}
 \mathrm{Tr}_E^{pn}\left[\left(\frac{\Omega^*_{cav}}{2}\Ket{0} \Bra{X} B_- + \frac{\Omega_{cav}}{2}\Ket{X} \Bra{0} B_+\right)\rho^{pn}_E\right] \nonumber \\
 = \frac{\Omega^*_{cav}}{2}\langle B \rangle\Ket{0} \Bra{X} + \frac{\Omega_{cav}}{2}\langle B \rangle\Ket{X} \Bra{0} ~,
 \end{align}
 where 
\begin{equation}
 \langle B \rangle =  \exp\left[ -\frac{1}{2}\int_0^\infty \mathrm{d} \omega \frac{J_{pn}(\omega)}{\omega^2} \coth(\beta \omega / 2) \right] ~. 
 \end{equation}
In order to expand perturbatively, we therefore define the system-bath interaction with respect to this value. To this end, we add the expectation value by defining $\mathcal{B}_\pm = B_\pm - \langle B \rangle$ and $\Omega^{pn}_{cav} = \langle B \rangle \Omega_{cav}$ and regrouping our system and interaction Hamiltonian terms, obtaining:
\begin{align}\label{newinteraction}
H_{SP} &= \delta' \Ket{X}\Bra{X} + \frac{\Omega^{pn *}_{cav}}{2}\Ket{0} \Bra{X} + \frac{\Omega^{pn}_{cav}}{2}\Ket{X} \Bra{0}~, \\
H_{IP}^{pn} &= \frac{\Omega^*_{cav}}{2}\Ket{0} \Bra{X} \mathcal{B}_- + \frac{\Omega_{cav}}{2}\Ket{X} \Bra{0} \mathcal{B}_+ ~,
\end{align}
As for Eq.~\eqref{eq:compactHpti}, we introduce operator labels $B^{pn}_{1/2} = \mathcal{B}_\mp$, $A^{pn}_1 =  \Omega^*_{cav} /2 \, \Ket{0} \Bra{X}$ and $A^{pn}_2 = A_1^{pn \dagger}$ to recast the above interaction Hamiltonian into the compact form
\begin{equation}
\label{eq:compactHphi}
H_{IP}^{pn} = \sum_{i=1}^2 A^{pn}_i \otimes B^{pn}_i 
\end{equation}
which will prove useful for the derivation of the master equation.

\subsection{Master Equation}\label{cavityME}

Having obtained our Hamiltonian in the polaron frame and partitioned it into system, interaction and environment parts, we can make use of the generically derived microscopic second-order Born-Markov master equation of Ref.~\cite{breuer} (Eqn. 3.118). The interaction terms Eqs.~\eqref{eq:compactHpti} and \eqref{eq:compactHphi} are of the required form underlying this derivation, and the resultant ME (in the interaction picture) reads: 
\begin{align}\label{generalME}
\diff{}{t} &\rho_{SP}(t) = \\
 &-\int_0^\infty \mathrm{d}\tau \; \mathrm{Tr}_E [ H_{IP}(t), [ H_{IP}(t-\tau), \rho_{SP}(t)\otimes\rho_E(0) ] ]~, \nonumber
\end{align}
where $H_{IP}(t) = H^{pn}_{IP}(t) + H^{pt}_{IP}(t)$, and $\mathrm{Tr}_E$ denotes the trace over both environments \cite{breuer}. It can be easily shown \cite{phononreview} that the right-handside (RHS) of the above equation can be split into two parts:

\begin{align}\label{splitgeneralME}
\diff{}{t} &\rho_{SP}(t) = \\
 &-\int_0^\infty \mathrm{d}\tau \mathrm{Tr}^{pn}_E [ H^{pn}_{IP}(t), [ H^{pn}_{IP}(t-\tau), \rho_{SP}(t)\otimes\rho^{pn}_E(0) ] ]~ \nonumber \\
 &-\int_0^\infty \mathrm{d}\tau \mathrm{Tr}_E [ H^{pt}_{IP}(t), [ H^{pt}_{IP}(t-\tau), \rho_{SP}(t)\otimes\rho_E(0) ] ]~. \nonumber
\end{align}
Since we assume that the (initial) environmental state is thermal, $\rho_E(0) $ factorises: $\rho_E(0) = \rho^{pn}_E(0) \otimes \rho^{pt}_E(0)$.

\subsubsection{Phonon bath correlations}

We proceed by analysing the first term on the RHS of Eq.~\eqref{splitgeneralME} which captures the influence of phonons on the TLS dynamcis with scattering rates determined by phonon correlation functions \cite{Ulhaq2013, PhononRates, spectrum}. In the ME formalism, the rate $\gamma(\omega)$ of a dissipative process is given by $\gamma(\omega) = 2 \mathrm{Re}\left[ \int_0^\infty \mathrm{d}s K(s) \right]$, where $K(s)$ is the relevant correlation function [{\it c.f.}~Eq.~(3.137) in Ref.~\cite{breuer}]. For our phonon dissipator, these functions are given by
\begin{align} 
C^{pn}_{ii}(\tau) &= \mathrm{Tr}^{pn}_{E} \left[ \mathcal{B}^\dagger_\pm(\tau) \mathcal{B}_\pm(0) \rho^{pn}_E(0)\right]  \nonumber\\ 
& = \langle B \rangle^2 (\mathrm{e}^{\phi(\tau)} -1)~, \label{phonon_cor_fns_cav:reg}\\
C^{pn}_{ij}(\tau) &= \mathrm{Tr}^{pn}_{E} \left[ \mathcal{B}^\dagger_\pm(\tau) \mathcal{B}_\mp(0) \rho^{pn}_E(0)\right] \nonumber \\
&= \langle B \rangle^2 (\mathrm{e}^{-\phi(\tau)} -1)~, \label{phonon_cor_fns_cav:cross}
\end{align}
where $i, j \in \{ 1,2 \}$, $i \neq j$. After some algebra, we obtain a phonon dissipator of the form
\begin{align*}
\begin{split}
&\gamma^{pn}(\omega') \mathcal{L}[\sigma_-]  + \gamma^{pn}(-\omega') \mathcal{L}[\sigma_+] \\[10pt]
&- \gamma^{pn}_{cd}(\omega')  \mathcal{L}_{cd}[\sigma_-] - \gamma^{pn}_{cd}(-\omega') \mathcal{L}_{cd}[\sigma_+]  ~,
\end{split}
\end{align*}
where $\mathcal{L}[C] = C \rho_{SP} C^\dagger - \frac{1}{2}\{C^\dagger C, \rho_{SP} \}$ and $\mathcal{L}_{cd}[C] = C \rho_{SP} C - \frac{1}{2}\{C^2, \rho_{SP} \}$. The rates $\gamma^{pn}(\pm\omega')$ and $\gamma_{cd}^{pn}$ are
\begin{align*} 
\gamma^{pn}(\pm \omega') &= \frac{\left| \Omega_{cav}^{pn} \right|^2}{4} \int_{-\infty}^\infty \mathrm{d}\tau \; \mathrm{e}^{\pm i \omega' \tau} \left( \mathrm{e}^{\phi(\tau)} - 1 \right)~, \\
\gamma^{pn}_{cd}(\omega') &= \frac{\left( \Omega^{pn*}_{cav} \right)^2}{4} \int_{-\infty}^\infty \mathrm{d}\tau \; \cos(\omega' \tau) \left( 1- \mathrm{e}^{-\phi(\tau)} \right)~, \\
\gamma^{pn}_{cd}(-\omega') &= \frac{\left( \Omega_{cav}^{pn} \right)^2}{4} \int_{-\infty}^\infty \mathrm{d}\tau \; \cos(\omega' \tau) \left( 1- \mathrm{e}^{-\phi(\tau)} \right)~,
\end{align*}
where $\phi(\tau) = \int_0^\infty \mathrm{d} \omega \frac{J_{pn}(\omega)}{\omega^2} [\coth(\beta \omega / 2)\cos(\omega \tau) - i \sin(\omega \tau)]$. Our rates match the ones obtained by Roy-Choudhury \textit{et al.} \cite{PhononRates} in previous work\footnote{Ref.~\cite{PhononRates} introduces an additional, phenomenological, pure dephasing term, which we have not included in this paper.}. The rates $\gamma^{pn}(\omega')$ and $\gamma^{pn}(-\omega')$ correspond to enhanced radiative decay and incoherent excitation of the TLS, respectively, whilst $\gamma^{pn}_{cd}(\pm\omega')$ is associated with cross-dephasing \cite{Ulhaq2013}.

\subsubsection{Electromagnetic bath correlations}\label{EM_bath_cor_cav}

Having arrived at a `Lindblad-like' phonon dissipator\footnote{Note that we have not performed a secularisation and our ME is therefore not strictly of Lindblad form}, we now turn our attention to the second term of the RHS of Eq.~\eqref{splitgeneralME}. This term will yield the modified SE rate of the TLS near the cavity, as well as account for the frequency shift via a unitary renormalisation term. As in the previous section, we begin by explicitly printing the correlation functions obtained from Eq.~\eqref{splitgeneralME}:
\begin{align}\label{photon_cor_fns_cav}
&C^{pt}_{ij}(\tau) \\
&= \mathrm{Tr}_{E} \left[ \left(B^{pt \dagger}_i(\tau) \otimes C^\dagger_i(\tau) \right) \left( B^{pt}_j(0) \otimes C_j(0) \right) \rho_E(0)\right]~, \nonumber \\
&= \mathrm{Tr}^{pn}_{E} \left[ B^{pt \dagger}_i(\tau) B^{pt}_j(0)  \rho^{pn}_E(0)\right]  \mathrm{Tr}^{pt}_{E} \left[ C^\dagger_i(\tau) C_j(0) \rho^{pt}_E(0)\right]~, \nonumber
\end{align}
where $i, j \in \{ 1,2 \}$. After substituting for the bath operators, we make use of the following relations \cite{breuer}
\begin{align*}
\mathrm{Tr}^{pt}_{E} \left[ a_{\mathbf{q} \lambda} a_{\mathbf{q}' \lambda'} \rho^{pt}_E(0) \right] &= \mathrm{Tr}^{pt}_{E} \left[ a^\dagger_{\mathbf{q} \lambda} a^\dagger_{\mathbf{q}' \lambda'} \rho^{pt}_E(0) \right] &=&~0 ~,\\
\mathrm{Tr}^{pt}_{E} \left[ a_{\mathbf{q} \lambda} a^\dagger_{\mathbf{q}' \lambda'} \rho^{pt}_E(0) \right]  &= \delta_{\mathbf{q}\mathbf{q}'}\delta_{\lambda \lambda'} (1 + N(\nu_\mathbf{q})) &\approx&~\delta_{\mathbf{q}\mathbf{q}'}\delta_{\lambda \lambda'}~, \\[10pt]
\mathrm{Tr}^{pt}_{E} \left[ a^\dagger_{\mathbf{q} \lambda} a_{\mathbf{q}' \lambda'} \rho^{pt}_E(0) \right]  &= \delta_{\mathbf{q}\mathbf{q}'}\delta_{\lambda \lambda'} N(\nu_\mathbf{q}) &\approx& ~0~, 
\end{align*}
where we have assumed that $\forall \omega > 0$, the Planck distribution $N(\omega) \approx 0$\footnote{Only (optical) photon modes with energies close to $\omega_0$ are relevant, for which this approximation is typically justified under ambient conditions. However, the generalisation to a finite temperature photon bath is also straightforward.}. This means that we only have a single non-vanishing correlation function $C^{pt}_{11}(\tau)$. Following Ref.~\cite{FermiGR2}, we consider well-separated photon and phonon correlation times (appropriate for an unstructured photonic environment), so that $C^{pt}_{11}(\tau)$ reduces to the photon bath correlation function in the absence of a phonon bath. The latter is given by 
\begin{equation}
C^{pt}_{11}(\tau) = \frac{|\mathbf{d}|^2 }{6 \pi^2 \epsilon c^3}\int_0^\infty \mathrm{d}\nu_\mathbf{q} \; \nu^3_\mathbf{q} [1 + \mathcal{F}_{cav}(q r_d)]~,
\end{equation}
where the term
\begin{figure*}[ht!]
\centering
  \begin{subfigure}[b]{0.5\textwidth}
    \begin{flushleft}
    \includegraphics[width=0.9\textwidth]{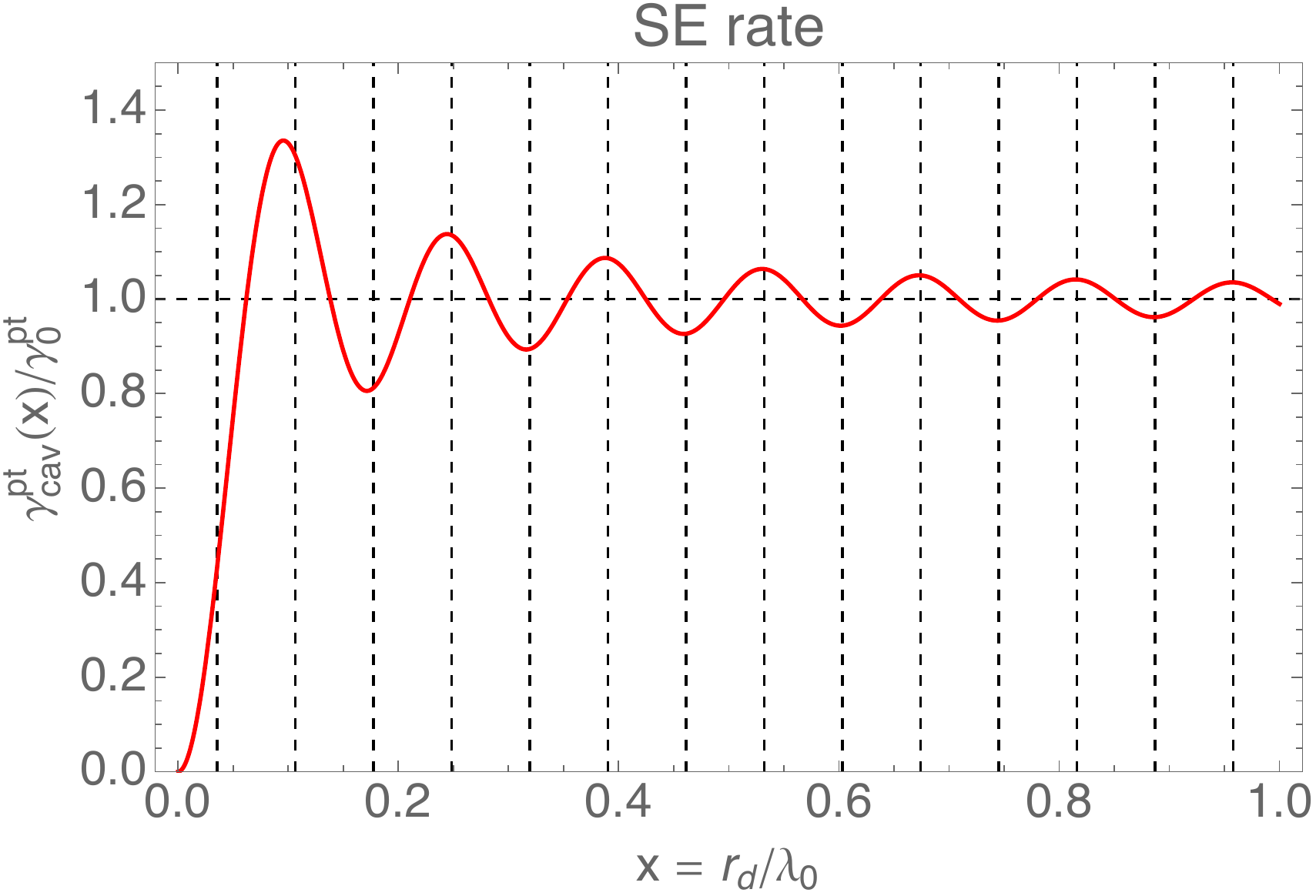} 
    \end{flushleft}
  \end{subfigure}
  \begin{subfigure}[b]{0.5\textwidth}
    \begin{flushleft}
    \includegraphics[width=0.9\textwidth]{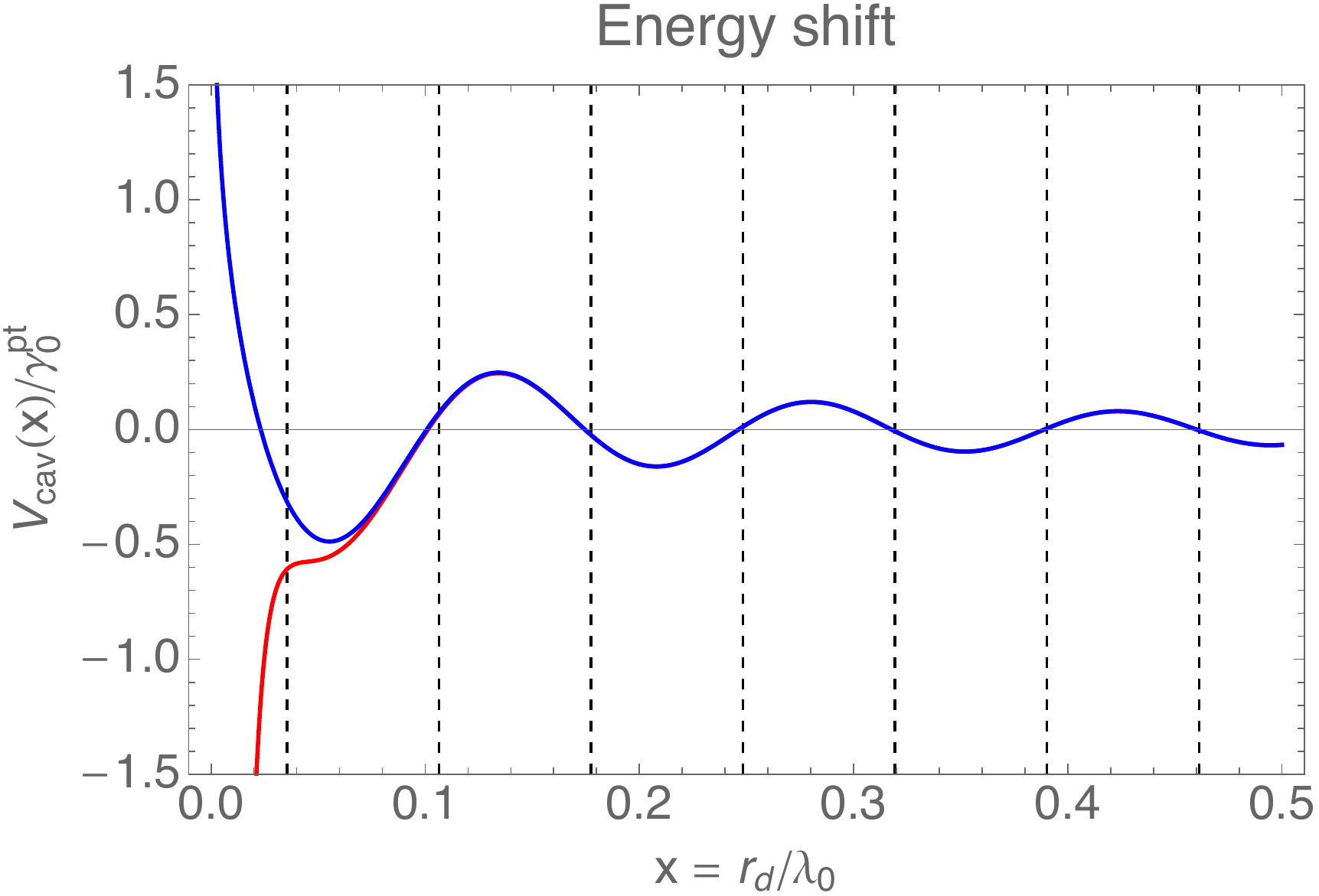} 
    \end{flushleft}
  \end{subfigure} 
  \caption{Spontaneous emission rate (left) and energy shift (right) for the half-sided cavity model (red), where we divided expressions \eqref{SE_rate_cav} and \eqref{energy_shift_cav} by the bare SE rate in order to avoid dependence on its value. The blue energy shift curve denotes the energy shift obtained using a full QED approach \cite{agarwal3}, showing a distinctively different behaviour at smaller separations ($\lesssim 0.05 \lambda_0$) when compared to the half-sided cavity and image approaches. The oscillations persist even at larger separations, of the order of the emission wavelength $\lambda_0$ for the SE rate. As $x\rightarrow\infty$, the SE rate tends to that of a bare emitter and the energy shift vanishes, as expected.}
  \label{energyrate}
\end{figure*}
\begin{equation}\label{Fcav}
\mathcal{F}_{cav}(x) = \frac{3}{2}\left( -\frac{\sin(2 x)}{2 x} - \frac{\cos(2 x)}{(2 x)^2} + \frac{\sin(2 x)}{(2 x)^3} \right)~,
\end{equation}
describes the influence of the metal surface.  The SE rate then evaluates to
\begin{equation}\label{SE_rate_cav}
\gamma_{cav}^{pt}(\omega') = (1+ \mathcal{F}_{cav}(q_0 r_d))\gamma_0^{pt}(\omega') ~,
\end{equation}
where $\gamma^{pt}_0(\omega')$ is the bare SE rate for an isolated TLS, and is given by $\gamma^{pt}_0(\omega') = |\mathbf{d}|^2 \omega'^3 / 3 \pi \epsilon c^3$. The imaginary part of the correlation tensor has two components: the first term is the usual Lamb shift (whose expression is divergent unless one adopts a full QED approach based on a relativistic Hamiltonian and appropriate renormalisation \cite{Gardiner}). The second term is the additional energy shift term and takes the form \cite{agarwal, agarwal2, ficek2005quantum}
\begin{equation}\label{energy_shift_cav}
V_{cav} = \frac{1}{2}  \mathcal{G}_{cav}(q_0 r_d)\gamma^{pt}_0(\omega')~,
\end{equation}
where the function $\mathcal{G}_{cav}$ is given by
\begin{equation}
\mathcal{G}_{cav}(x) = \frac{3}{2} \left(- \frac{\sin(2 x)}{(2 x)^2} - \frac{\cos(2 x)}{(2 x)^3}  + \frac{\cos(2 x)}{2 x}\right)~.
\end{equation}
Overall, the transition frequency for the TLS in the polaron frame is now given by
\begin{equation}\label{shifted_cavity_frequency}
\tilde{\omega}' = \omega' + V_{cav}~
\end{equation}
and the final polaron frame ME  takes the following form in the Schr\"odinger picture:
\begin{align}\label{ME_polaron_cav}
\begin{split}
\diff{}{t} &\rho_{SP} =  \\
             &-\frac{i}{\hbar} [H'_{SP}, \rho_{SP}(t)] + D_{pn}(\rho_{SP}) +D_{pt}(\rho_{SP}) ~,
\end{split}
\end{align}
where $D_{pn}(\rho_{SP}) = \gamma^{pn}(\omega') \mathcal{L}[\sigma_-]  + \gamma^{pn}(-\omega') \mathcal{L}[\sigma_+] - \gamma^{pn}_{cd}(\omega')  \mathcal{L}_{cd}[\sigma_-] - \gamma^{pn}_{cd}(-\omega') \mathcal{L}_{cd}[\sigma_+]  $ and $D_{pt}(\rho_{SP}) = \gamma^{pt}_{cav}(\omega') \mathcal{L}[\sigma_-]$. $H'_{SP}$ is the system Hamiltonian in the polaron frame including the energy shift from Eq.~\eqref{energy_shift_cav}.

In summary, Eqs.~\eqref{SE_rate_cav} and \eqref{energy_shift_cav} capture how the presence of a metal surface (here treated as a perfect reflector) alters the SE rate and the transition frequency of the TLS, respectively. 
Considering our results in the absence of phonons, we find full analytical agreement with the prior literature on the image dipole approach \cite{ficek2005quantum,agarwal3}, and except for very small separations, we also have excellent numerical agreement with the full QED approach \cite{agarwal3}. We show this agreement in Fig.~\ref{energyrate} as a function of the distance of the emitter to the surface. The dashed vertical lines at multiples of $1/8 n$ (where $n$ is the refractive index of the host material, taken to be GaAs in our case), taken from Eqns.~\eqref{SE_rate_cav} and \eqref{energy_shift_cav}, serve as a guide to the eye for the approximate frequency of oscillation, and demonstrate that multiple periods occur within a wavelength's separation of emitter to surface. In the limiting case $r_d \rightarrow \infty$, we have $V_{cav} \rightarrow 0$ and $\gamma_{cav}^{pt}(\omega') \rightarrow \gamma_0^{pt}(\omega')$, i.e.~we recover the case of an isolated QD as required.

\section{Image Emitter Approach}\label{image:dot:approach}

Models involving emission from a combination of two identical TLS have been used extensively to study the modifications to the SE rate of an emitter in the proximity of a dielectric or metal surface. After setting up the appropriate Hamiltonian, we shall once more derive a polaron frame ME. We then show that this ME is identical to the one derived using the half-sided cavity approach, provided we disregard certain terms in order to constrain the dynamics of our two emitter model to the `right' subspace. 

\subsection{Setup}

We focus on the case where the dipole is oriented parallel to the surface\footnote{We discuss modifications for the perpendicular case in the Appendix} (as is appropriate for a typical self-assembled QD emitter), implying that the image dipole will be antiparallel \citep{babiker,agarwal,agarwal2,agarwal3}. In what follows, we shall once again take the {\it real} emitter to be situated at a distance $r_d > 0$ along the positive $z$-axis, with the dipole vector oriented in the positive $x$-direction. Hence, the corresponding {\it image} dipole is positioned at $z = -r_d$, with its dipole vector being parallel to the negative $x$-axis. 

\subsection{Hamiltonian}

The Hamiltonian of the two driven TLS in a frame rotating with frequency $\omega_l$ is given by
\begin{equation}
H_S = \sum_{j=1}^2 \delta \Ket{X_j}\Bra{X_j} + \frac{\Omega^*_j}{2}\Ket{0_j} \Bra{X_j} + \frac{\Omega_j}{2}\Ket{X_j} \Bra{0_j} ~,
\end{equation}
where the subscript $j=1, 2$ denotes the real and image TLS, respectively. In order to match the boundary conditions required for reflection, we model the classical driving field as two counter-propagating beams, with the secondary `reflected' beam having a $\pi$ phase shift with respect to the original beam. For simplicity, we model these as plane waves propagating along the $z$-axis and polarised in the $x$-direction. In phasor notation, these two waves can be written as
\begin{align}
\begin{split}
\mathbf{E}_1(\mathbf{r}) &= \mathbf{E}_{incident}(\mathbf{r}) = E_0 \mathrm{e}^{i \mathbf{q}_l \cdot \mathbf{r}} \hat{\mathbf{x}} ~, \\
\mathbf{E}_2(\mathbf{r}) &= \mathbf{E}_{reflected}(\mathbf{r}) = -E_0 \mathrm{e}^{-i \mathbf{q}_l \cdot \mathbf{r}} \hat{\mathbf{x}}~,
\end{split}
\end{align}
giving rise to the following Rabi frequencies at the positions $\mathbf{r}_{1,2}$ of the two emitters:
\begin{align}
\begin{split}
\Omega_1 &= 2 \mathbf{d}_1 \cdot (\mathbf{E}_1(\mathbf{r}_1) + \mathbf{E}_2(\mathbf{r}_1))~, \\
\Omega_2 &= 2 \mathbf{d}_2 \cdot (\mathbf{E}_1(\mathbf{r}_2) + \mathbf{E}_2(\mathbf{r}_2)) ~.
\end{split}
\end{align}
Since $\mathbf{r}_2 = -\mathbf{r}_1$ and $\mathbf{d}_2 = -\mathbf{d}_1$, we have $\Omega \coloneqq \Omega_1 = \Omega_2$.

We now turn to the wider electromagnetic environment (excluding the coherent driving field discussed above). The electric field operator can be written as in Eq.~\eqref{electricfield} but with the spatial mode functions now being replaced by the free-space functions
\begin{equation}
\mathbf{u}_{\mathbf{q} \lambda}(\mathbf{r}) = \sqrt{\frac{\omega_{\mathbf{q} \lambda}}{2 \epsilon V}} \mathbf{e}_{\mathbf{q} \lambda} \mathrm{e}^{i \mathbf{q} r} ~.
\end{equation}
The interaction Hamiltonian of the TLS with the photonic environment is then given by
\begin{align}
\begin{split}
H^{pt}_I =&H^{pt,1}_I + H^{pt,2}_I \\
              =&-\sum_{j=1}^2 \mathbf{d}_j \cdot \mathbf{E}(\mathbf{r_j}) (\Ket{0_j} \Bra{X_j} + \Ket{X_j} \Bra{0_j})~.
\end{split}
\end{align}

For the interaction with vibrational modes, we assume that both real and image TLS see the same phonon bath and possess perfectly correlated coupling constants $g_\mathbf{k}$. This  ensures the image system exactly follows the dynamics of real dipole, as is required for matching the boundary condition of a perfectly reflecting interface. Thus, our relevant Hamiltonian reads
\begin{align}
\begin{split}
H^{pn}_I =&H^{pn,1}_I + H^{pn,2}_I \\
	      =&\sum_{j=1}^2 \sum_{\mathbf{k}} \Ket{X_j} \Bra{X_j} g_\mathbf{k} ( b^\dagger_\mathbf{k} + b_\mathbf{k} )~.
\end{split}
\end{align}
Next, we move into the polaron frame with the transformation $\mathrm{e}^{S_1+S_2} = \mathrm{e}^{S_1}\mathrm{e}^{S_2}$, obtaining the transformed Hamiltonians
\begin{align}
H_{SP} = & \sum_{j=1}^2 \delta' \Ket{X_j}\Bra{X_j} + \frac{\Omega^{pn *}}{2}\Ket{0_j} \Bra{X_j} + \mathrm{H.c.}~, \\
H^{pt,j}_{IP} = &i \Ket{0_j} \Bra{X_j} B_-  \sum_{\mathbf{q}, \lambda} \mathbf{d}_j\cdot\mathbf{u}^*_{\mathbf{q}\lambda}(\mathbf{r}_j) a^\dagger_{\mathbf{q}\lambda} \nonumber \\
                        -&i \Ket{X_j} \Bra{0_j} B_+  \sum_{\mathbf{q}, \lambda} \mathbf{d}_j\cdot\mathbf{u}_{\mathbf{q}\lambda}(\mathbf{r}_j) a_{\mathbf{q}\lambda} ~,\nonumber  \\
H^{pn,j}_{IP} = &\frac{\Omega^*}{2}\Ket{0_j} \Bra{X_j} \mathcal{B}_- + \frac{\Omega}{2}\Ket{X_j} \Bra{0_j} \mathcal{B}_+ ~.
\end{align} 
As in Sec.~\ref{halfcavity}, the latter two can easily be seen to be of the following generic form (with appropriate identifications for the $A, B, C$ operators) which will enable straightforward use of the ME (3.118) from Ref.~\cite{breuer}:
\begin{align}                      
H^{pn,j}_{IP} = &\sum_{i=1}^2 A^{pn,j}_i \otimes B^{pn,j}_i~, \\
H^{pt,j}_{IP} = &\sum_{i=1}^2 A^{pt,j}_i \otimes B^{pt,j}_i \otimes C^j_i ~.
\end{align}  
\subsection{Master equation}

The ME for our system can, once again, be written as
\begin{align}\label{splitgeneralME2}
\diff{}{t} &\rho_{SP}(t) = \\
 &-\int_0^\infty \mathrm{d}\tau \mathrm{Tr}^{pn}_E [ H^{pn}_{IP}(t), [ H^{pn}_{IP}(t-\tau), \rho_{SP}(t)\otimes\rho^{pn}_E(0) ] ]~ \nonumber \\
 &-\int_0^\infty \mathrm{d}\tau \mathrm{Tr}_E [ H^{pt}_{IP}(t), [ H^{pt}_{IP}(t-\tau), \rho_{SP}(t)\otimes\rho_E(0) ] ]~, \nonumber
\end{align}
however, it now features a larger number of correlation functions due to the presence of the image emitter. Following the general procedure in Sec.~\ref{cavityME}, we shall analyse different contributions in turn to arrive at our final ME of the image emitter model.

\subsubsection{Phonon dissipator}

The correlation functions (including cross correlation terms between bath operators of the real and image system) result in the following phonon dissipator
\begin{align}
D_{pn}&(\rho_{SP})= \\
&\sum_{i,j=1}^2 \gamma^{pn}_{ji}(\omega') \left( \sigma^j_- \rho_{SP}(t) \sigma^i_+ - \frac{1}{2}\{ \sigma^i_+ \sigma^j_-, \rho_{SP}(t) \} \right) \nonumber \\
+&\sum_{i,j=1}^2 \gamma^{pn}_{ji}(-\omega') \left( \sigma^j_+ \rho_{SP}(t) \sigma^i_- - \frac{1}{2}\{ \sigma^i_- \sigma^j_+, \rho_{SP}(t) \} \right) \nonumber \\
-&\sum_{i,j=1}^2 \gamma^{pn}_{cd, ji}(\omega') \left( \sigma^j_- \rho_{SP}(t) \sigma^i_- - \frac{1}{2}\{ \sigma^i_- \sigma^j_-, \rho_{SP}(t) \} \right) \nonumber \\
-&\sum_{i,j=1}^2 \gamma^{pn}_{cd, ji}(-\omega') \left( \sigma^j_+ \rho_{SP}(t) \sigma^i_+ - \frac{1}{2}\{ \sigma^i_+ \sigma^j_+, \rho_{SP}(t) \} \right)~,\nonumber 
\end{align}
where the rates $\gamma^{pn}_{ji}(\pm\omega')$ and $\gamma^{pn}_{cd, j}$ are given by
\begin{align*}\label{phononratesRealandImage}
\gamma^{pn}_{ji}(\pm \omega') &= \frac{|\Omega^{pn}|^2}{4} \int_{-\infty}^\infty \mathrm{d}\tau \; \mathrm{e}^{\pm i \omega' \tau} \left( \mathrm{e}^{\phi(\tau)} - 1 \right)~, \\
\gamma^{pn}_{cd, ji}(\omega') &= \frac{ (\Omega^{pn*})^2}{4} \int_{-\infty}^\infty \mathrm{d}\tau \; \cos(\omega' t) \left( 1- \mathrm{e}^{-\phi(\tau)}  \right)~, \\
\gamma^{pn}_{cd, ji}(-\omega') &= \frac{ (\Omega^{pn})^2}{4} \int_{-\infty}^\infty \mathrm{d}\tau \; \cos(\omega' t) \left( 1- \mathrm{e}^{-\phi(\tau)}  \right) ~.
\end{align*}
We shall return back to the phonon dissipator when discussing the ME equation in the symmetric-antisymmetric basis, which allows us to derive a model agreeing with the half-sided cavity approach.

\subsubsection{Photon dissipator}

We now turn our attention to the photon dissipator term from Eq.~\eqref{splitgeneralME2}. After evaluating the correlation and cross-correlation functions, we obtain the usual expression for two emitters \cite{ficek2005quantum} in a shared electromagnetic environment,
\begin{align}
\begin{split}
D_{pt}&(\rho_{SP})= \\
&\sum_{i,j=1}^2 \gamma^{pt}_{ji} \left( \sigma^j_- \rho_{SP}(t) \sigma^i_+ - \frac{1}{2}\{ \sigma^i_+ \sigma^j_-, \rho_{SP}(t) \} \right) ~,
\end{split}
\end{align}
where the diagonal terms $\gamma^{pt}_{22}(\omega') = \gamma^{pt}_{11}(\omega') =  \gamma_0^{pt}(\omega')$, whilst the off diagonal terms are given by $\gamma^{pt}_{12}(\omega') = \gamma^{pt}_{21}(\omega') =  \mathcal{F}_{12}(q_0 \Delta r)\gamma_0^{pt}(\omega')$ with $\Delta r = r_1 - r_2 = 2r_d$, and where
\begin{align}\label{Fimage}
\begin{split}
\mathcal{F}_{12}(x) = \frac{3}{2}\left( -\frac{\sin(x)}{x} - \frac{\cos(x)}{x^2} + \frac{\sin(x)}{x^3} \right)~.
\end{split}
\end{align}
This is the same function obtained for the half-sided cavity approach [{\it c.f.}~Eq.~\eqref{Fcav}]. The imaginary part of the correlation function yields the `correction' term to the unitary part of the ME \cite{breuer,ficek2005quantum,agarwal}: its diagonal contribution represents diagonal Lamb shift terms. Their small energetic shifts can be absorbed into the bare TLS transition frequency. We thus focus on the off-diagonal element which is of the form:
\begin{equation}\label{energy_shift_image}
V_{12} = \frac{1}{2}  \mathcal{G}_{12}(q \Delta r)\gamma^{pt}_0(\omega')~,
\end{equation}
where the function $\mathcal{G}_{12}$ is
\begin{align}
\begin{split}
\mathcal{G}_{12}(x) = \frac{3}{2} \left(- \frac{\sin(x)}{x^2} - \frac{\cos(x)}{x^3}  + \frac{\cos(x)}{x}\right)~.
\end{split}
\end{align}
Again, this corresponds to the same energy shift term we have previously encountered in Sec.~\ref{EM_bath_cor_cav}. After diagonalising the Hamiltonian, the frequency of the symmetric excited to ground state transition (in the polaron frame) is then given by 
\begin{equation}
\tilde{\omega}' = \omega' + V_{12}~,
\end{equation}
exactly matching the transition frequency Eq.~\eqref{shifted_cavity_frequency} of the half-sided cavity model.

\begin{figure}[ht!]
\centering
\def\svgwidth{0.9\columnwidth}
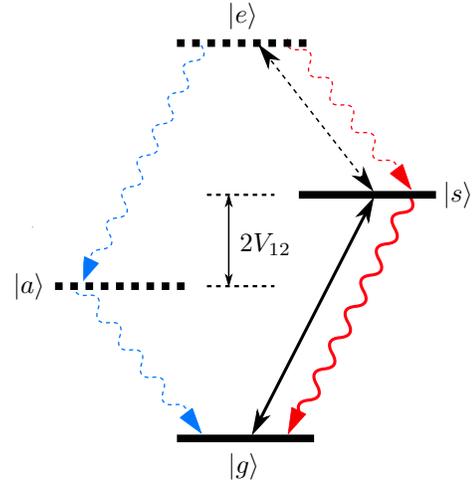
\caption{\label{lvl_diag}  Energy level diagram for the two emitter system. The symmetric ($\Ket{s}$) and antisymmetric ($\Ket{a}$) levels are shifted up and down by $V_{12}$, respectively. The black arrows indicate the laser driving; the antisymmetric state is decoupled. Blue and red wavy lines indicate photon emission from the antisymmetric and symmetric channel, respectively. As discussed in the text, it is necessary to disable driving on the $\Ket{s} \leftrightarrow \Ket{e}$ transition (black dashed) to recover the effective two level-system $\Ket{g} \leftrightarrow \Ket{s}$. For environments permitting photon absorption, the dashed wavy transitions also need to be explicitly disabled. }
\end{figure}

\subsection{Effective TLS in the energy eigenbasis}\label{eigbasis}

\begin{figure*}[ht!]
\begin{flushleft}
\def\svgwidth{1\textwidth}
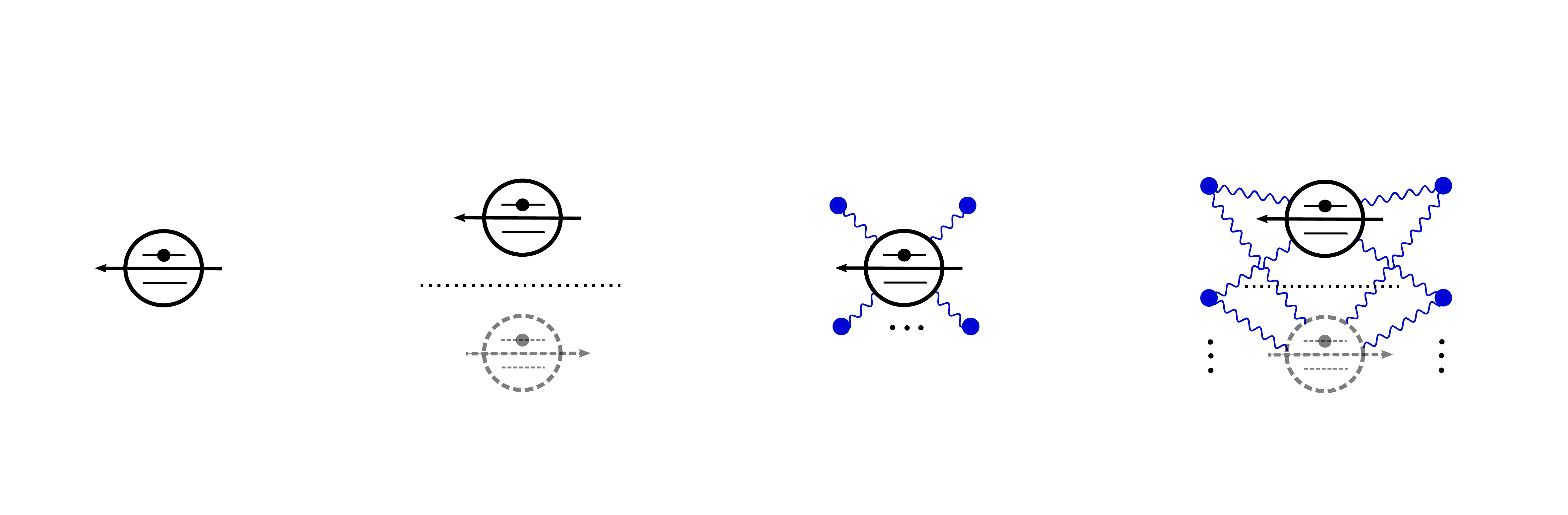
\caption{\label{schematic2}Overview of the four scenarios for an optical dipole considered in this work.  All cases have a schematic depiction accompanied by the corresponding SE rates $\gamma_0$ and transition frequencies $\omega$. Here, $\Delta r$ is the separation between the real and image dipole, $\mathcal{F}_{12}(q_0 \Delta r)$ and $V_{12}$ are given by Eqns.~\eqref{Fimage} and \eqref{energy_shift_image}, respectively, and $\omega_0$ and $\omega'$ are the bare and polaron shifted frequencies, respectively. The blue `masses on springs' (blue circles) denote the phonon bath. Note that the driving field is not shown here, as its presence or absence does not influencing the relevant properties.}
\end{flushleft}
\end{figure*}

As stated in the introduction, previous literature treating spontaneous emission from initially excited emitters considered the transition from the symmetrically excited to the ground state, as this choice yields matching results with other methods \cite{ficek2005quantum, babiker}. We follow this approach and adopt the basis $\{\Ket{e}, \Ket{s}, \Ket{a},\Ket{g}\}$ with $\Ket{s} = (\Ket{0_1} \Ket{X_2} + \Ket{X_1}\Ket{0_2}) / \sqrt{2}$ and $\Ket{a} = (\Ket{0_1} \Ket{X_2} - \Ket{X_1}\Ket{0_2}) / \sqrt{2}$, see Fig.~\ref{lvl_diag}. In this basis, our full polaron ME reads:
\begin{align}\label{fullME}
\begin{split}
\diff{}{t} \rho_{SP}(t) = &-\frac{i}{\hbar} [H'_{SP}, \rho_{SP}(t)]  \\
                                   &+ D^{s}_{pn}(\rho_{SP}) +D^{a}_{pt}(\rho_{SP})+D^{s}_{pt}(\rho_{SP}) ~,
\end{split}
\end{align}
where the dissipator terms are explicitly given in Appendix \ref{app:dissipators}. Here, $H'_{SP}$ denotes the system diagonalised Hamiltonian [including the energy shift term Eq.~\eqref{energy_shift_image}]. The ME photonic dissipator separates into a symmetric channel ($\Ket{g} \leftrightarrow \Ket{s} \leftrightarrow \Ket{e}$) and an antisymmetric one ($\Ket{g} \leftrightarrow \Ket{a} \leftrightarrow \Ket{e}$). Courtesy of the fully correlated phonon bath, phonons also only act in the symmetric channel.

Since $\Omega_1 = \Omega_2$, the symmetric channel Rabi frequency becomes $\Omega_{sg} \coloneqq (\Omega_1 + \Omega_2)/\sqrt{2} = \sqrt{2}\Omega = \Omega_{cav}$ and hence we obtain the same phonon rates as in the half-sided cavity approach\footnote{The last equality holds due to the difference in density of modes appearing in the derivation of the Rabi frequency in both models.}. Furthermore, the antisymmetric channel Rabi frequency $\Omega_{a} \coloneqq(\Omega_1 - \Omega_2)/\sqrt{2} = 0$, meaning that the laser field is completely decoupled from the antisymmetric state. 

Consistency with the Green's function and half-sided cavity approach demands that we restrict the dynamics of our four-dimensional Hilbert space to the subspace spanned by the states $\{ \Ket{g}, \Ket{s} \}$, i.e.~the larger Hilbert space only served to let us calculate the correct properties of this single transition. Fully decoupling the antisymmetric singly and the doubly excited states from the dynamics is achieved by disabling the laser driving on the $\Ket{s} \leftrightarrow \Ket{e}$ transition. For finite temperature photon environments with $N(\omega) \neq 0$, we also need to remove dissipative photon absorption channels, by dropping the  antisymmetric dissipator term $D^{a}_{pt}(\rho_{SP})$ from the ME and explicitly removing the dissipative $\Ket{s} \leftrightarrow \Ket{e}$ operator.

The image approach can thus be reduced to an effective TLS model featuring the same Rabi frequency, SE rate, and transition frequency as the half-sided cavity approach -- i.e.~ displaying full equivalence between the two representations. 
 
In Fig.~\ref{schematic2}, we summarise the key results from the previous sections: We show the transition frequency and SE rate for the all four cases considered in this Article alongside their schematic depictions. The driving term is not included as it has no direct influence on the properties of the optical dipole transition.

\section{Resonance Fluorescence Spectrum}

\begin{figure*}[t]
\centering
  \begin{subfigure}[b]{0.5\textwidth}
    \includegraphics[width=1.05\textwidth]{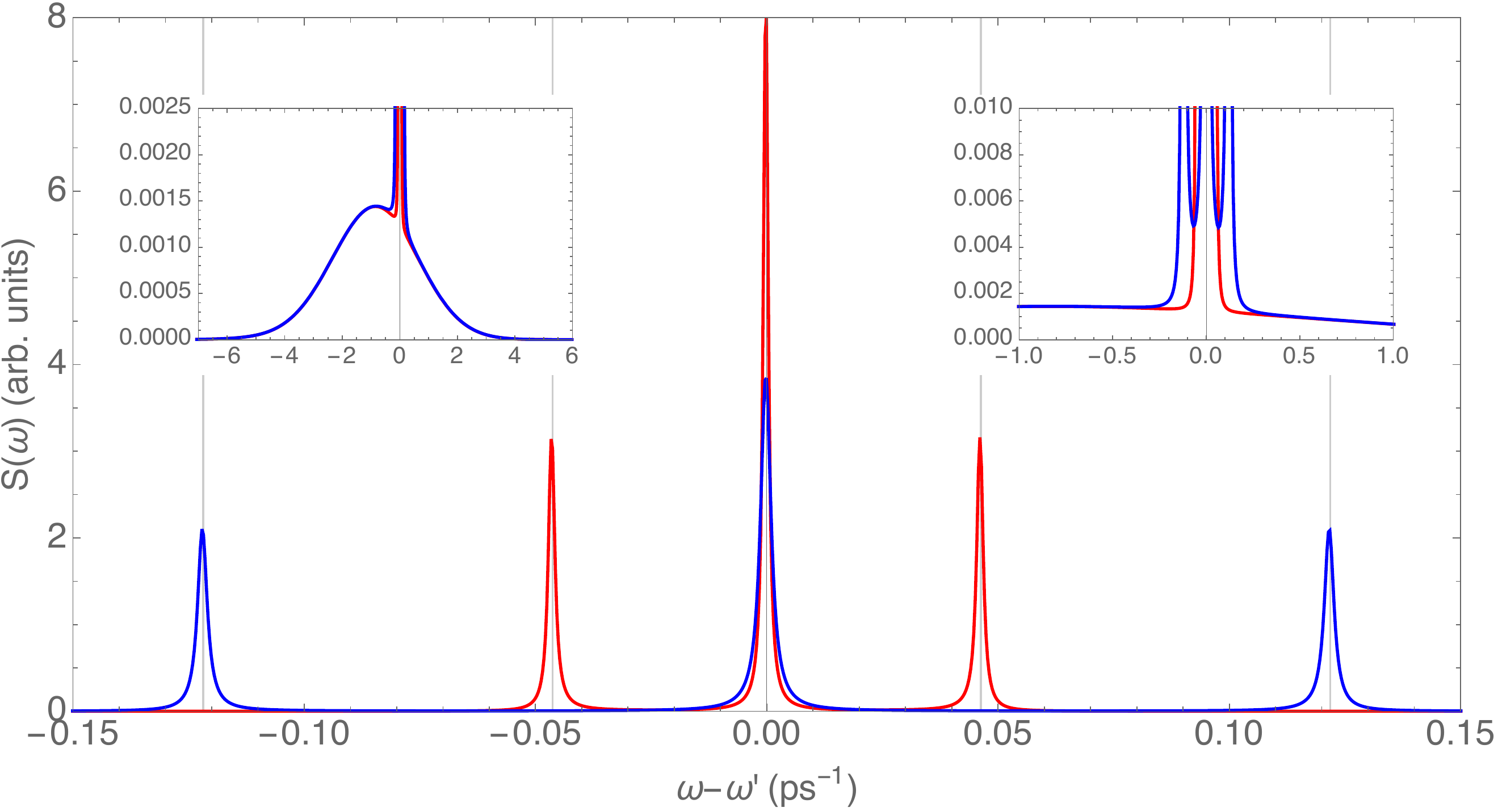} 
  \end{subfigure}
  \begin{subfigure}[b]{0.5\textwidth}
    \includegraphics[width=0.85\textwidth]{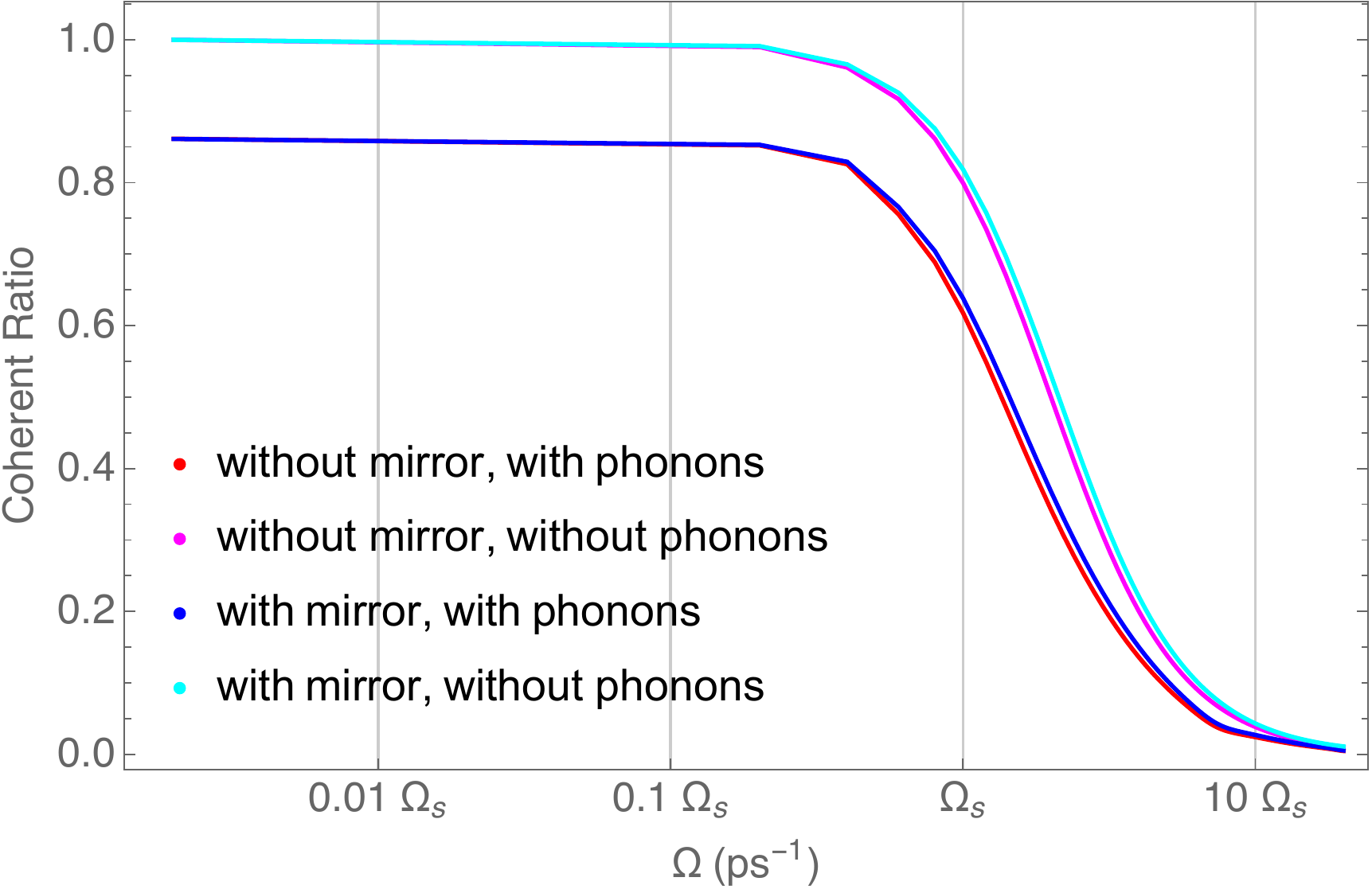} 
  \end{subfigure} 
  \caption{ {\bf Left:} Incoherent component of the RF spectrum for a single TLS (blue) and the effective TLS incorporating surface-induced modifications (red).  
  {\bf Right:} Ratio of coherent emission for all four cases (with/without mirror, with/without the phonon environment) as a function of the (normalised) effective Rabi frequency. $\Omega_s$ denotes the saturation Rabi frequency for $\gamma^{pt}_0 = 0.001$~ps$^{-1}$. See text for a discussion.}
  \label{spectrumplot}
\end{figure*}
Having included the possibility of laser driving in our model, a natural application is to study the resonance fluorescence (RF) spectrum of a condensed matter TLS near a mirroring surface. We use the ME \eqref{fullME} (after discarding the antisymmetric channel, as argued above) to calculate the spectral function, which is given by the Fourier transform of the (steady-state) first order correlation function $\mathrm{lim}_{t \rightarrow \infty}\langle \mathbf{E}^{(-)}(\mathbf{R}, t) \mathbf{E}^{(+)}(\mathbf{R}, t + \tau) \rangle$, where $\mathbf{E}^{(-)}(\mathbf{R}, t)$ and $\mathbf{E}^{(+)}(\mathbf{R}, t)$ are, respectively, the negative and positive components of the electric field operator evaluated at the position $\mathbf{R}$ of the detector \cite{ficek2005quantum}. These operators are related to the system operators $\sigma_- = \Ket{0}\Bra{X}$ and $\sigma_+ = \Ket{X}\Bra{0}$, and hence, after applying the polaron transformation, the RF spectral function can be written as
\begin{align}\label{spectrum}
\begin{split}
S(\omega) \propto \int_{-\infty}^\infty \mathrm{d}\tau &\mathrm{e}^{-i (\omega - \omega') \tau} \times \\
&\langle \sigma_+(\tau) B_+(\tau) \sigma_-(0) B_-(0) \rangle_{s}~, \\
\end{split}
\end{align}
where we have exploited the temporal homogeneity of the stationary correlation function, and where the subscript `s' denotes the trace taken with respect the steady-state density matrix \cite{breuer}. The correlation function appearing in Eq.~\eqref{spectrum} involves two timescales, the nanosecond timescale associated with the exciton lifetime, and the shorter picosecond phonon bath relaxation timescale, allowing us to separate the correlation function into the product $\langle \sigma_+(\tau) \sigma_-(0) \rangle_s \langle B_+(\tau) B_-(0) \rangle_{s}$ \cite{jake}. Substituting the expression for the phonon bath correlation function, we obtain the spectral function
\begin{align}\label{spectrum_simplified}
\begin{split}
S(\omega) \propto \langle B \rangle^2\int_{-\infty}^\infty \mathrm{d}\tau &\mathrm{e}^{-i (\omega - \omega') \tau} \times  \\ 
&\mathrm{e}^{\phi(\tau)} \langle \sigma_+(\tau) \sigma_-(0) \rangle_{s}~. \\
\end{split}
\end{align}
In the left panel of Fig.~\ref{spectrumplot}, we show the incoherent part of the emission spectrum of our surface-modified system as well as that of a reference TLS (also subject to the same phonon environment). Following Ref.~\cite{gerardot}, we take the TLS's position relative to the surface as $r_d \sim 177$ nm. The reference TLS is driven with `free space' Rabi frequency given by $\Omega^{pn} = 2 \langle B \rangle \mathbf{d}\cdot\mathbf{E}_0$. As expected, the curves differ in the position of the Mollow sidebands and the width of the three peaks, since the former is determined by the effective Rabi frequency and the later depends on the emission rate, which both undergo a change in the presence of a reflective surface. The two insets in the left panel of Fig.~\ref{spectrumplot} show the much broader phonon sideband, which receives  $\sim 16\%$ of the scattered photons for the chosen spectral density at a phonon temperature of T=10~K.

In the right panel of Fig.~\ref{spectrumplot}, we plot the fraction of coherently scattered photons as a function of the renormalised effective Rabi frequency. This ratio is obtained numerically as the (integrated) coherent spectrum divided by the total integrated spectrum. There are two pairs of curves: one with and one without phonons. For the former, the finite area under the phonon sideband means that the coherent fraction does not go to unity even when driving far below saturation. The level at which this fraction plateaus is phonon coupling strength and temperature dependent \cite{jake}. By contrast, in the absence of phonons, almost all light is coherently scattered at weak enough driving. The close agreement between the two curves in each pair bears testament to the fact that the surface-modified emitter largely behaves like a bare emitter once the effective Rabi frequency has been corrected for (with the slight remaining discrepancy due to modifications of the natural lifetime). Indeed, plotting this ratio directly as a function of the laser driving field amplitude reveals sizeable horizontal shifts between these two curves in each pair (not shown). 

\section{Summary and Discussion}

We have extended the method of images -- traditionally developed for capturing spontaneous emission in atomic ensembles near reflective interfaces --  to the case of a driven solid-state emitter near a metal surface. We have developed two approaches: a half-sided cavity and image dipole, and shown that the latter agrees with the former, but only when additional `selection rules' are introduced to constrain the dynamics to the relevant subspace. Both our approaches agree with a Green's function treatment in the absence of a vibrational environment. Through a rigorous derivation, we find that the emitter can indeed still be described as an effective (phonon-dressed) two-level system with appropriately modified properties, even in the presence of a phonon bath and for a driven system. Our calculated RF spectrum corroborates this observation.

We note that image dipole approach not only necessitates a larger Hilbert space but also involved a more cumbersome ME derivation than the half-sided cavity approach. This begs the questions whether such an image approach remains useful. We submit that the method of images can more easily accommodate larger numbers of emitters near a surface (of varying separation to the surface), as the problem then straightforwardly maps onto the case of several optical dipoles in a shared (free space) electromagnetic environment -- a problem which has been studied extensively, see, e.g.,~Ref.~\cite{ficek2005quantum}. Future work might investigate the role of geometry in configurations with $N>1$ emitters, possibly resulting in the enhancement of Dicke superradiance of an ensemble of solid state emitters \cite{sanders, Machnikowski:Superradiance}, or the use of mirrors to bring about other collective effects in the light matter interaction, for example inspired by a recent proposal for engineering the quantum-enhanced absorption of light \cite{superabsorption} or by harnessing sub-radiant collective states \cite{Scully2015, Higgins2015}.
 
Another interesting avenue for future work might be the study of charged quantum dots featuring excited trion states. In addition to the optical dipole, the image approach would then feature a separate permanent dipole. To a first approximation, we would expect this second dipole to be static, meaning it would not radiate and only modify the spectrum via energetic shifts. However, one might speculate whether the Coulomb interaction of the three charges involved in the trion state could slightly `wiggle' this dipole, making some radiative contribution to the overall spectrum conceivable.

\section*{Acknowledgements}

The authors thank Peter Kirton, Fabio Biancalana, and David Gershoni for valuable suggestions. D.S. thanks SUPA for financial support, T.S. acknowledges studentship funding from EPSRC under grant no  EP/G03673X/1, B. D. G. thanks the Royal Society, and E. M. G. acknowledges support from the Royal Society of Edinburgh and the Scottish Government.
 
\section*{Appendix}
\appendix
\section{Eigenbasis Dissipators}
\label{app:dissipators}

The dissipators of Sec.~\ref{eigbasis} are given by
\begin{align}
\begin{split}
&D^{s}_{pn}(\rho_{SP})\\&= 2 \gamma^{pn}(\omega') \Big[ (S_{se}+S_{gs}) \rho_{SP}(t) (S_{es}+S_{sg})\\ 
                                       &- \frac{1}{2}\{ (S_{ee}+S_{ss}), \rho_{SP}(t) \} \Big] \\[15pt]
                                       &+ 2 \gamma^{pn}(-\omega') \Big[ (S_{es}+S_{sg}) \rho_{SP}(t) (S_{se}+S_{gs})\\ 
                                       &- \frac{1}{2}\{ (S_{gg}+S_{ss}), \rho_{SP}(t) \} \Big] \\[15pt]
                                       &- 2 \gamma^{pn}_{cd}(\omega') (S_{se}+S_{gs}) \rho_{SP}(t)(S_{se}+S_{gs}) \\[10pt]
                                       &- 2 \gamma^{pn}_{cd}(-\omega') (S_{es}+S_{sg}) \rho_{SP}(t) (S_{es}+S_{sg})~,
\end{split}
\end{align}
\begin{align}
\begin{split}
&D^{a}_{pt}(\rho_{SP})\\&= 2 \gamma^{pt}(\omega') \Big[ (S_{se}+S_{gs}) \rho_{SP} (t)(S_{es}+S_{sg})\\ 
                                       &- \frac{1}{2}\{ (S_{ee}+S_{ss}), \rho_{SP}(t) \} \Big] \\[15pt]
                                       &+ 2 \gamma^{pt}(-\omega') \Big[ (S_{es}+S_{sg}) \rho_{SP}(t) (S_{se}+S_{gs})\\ 
                                       &- \frac{1}{2}\{ (S_{gg}+S_{ss}), \rho_{SP}(t) \} \Big]                                      
\end{split}
\end{align}
\begin{align}
\begin{split}
&D^{s}_{pt}(\rho_{SP})\\&= 2 \gamma^{pt}(\omega') \Big[ (S_{se}+S_{gs}) \rho_{SP}(t) (S_{es}+S_{sg})\\ 
                                       &- \frac{1}{2}\{ (S_{ee}+S_{ss}), \rho_{SP}(t) \} \Big] \\[15pt]
                                       &+ 2 \gamma^{pt}(-\omega') \Big[ (S_{es}+S_{sg}) \rho_{SP}(t) (S_{se}+S_{gs})\\ 
                                       &- \frac{1}{2}\{ (S_{gg}+S_{ss}), \rho_{SP} (t)\} \Big]~,
\end{split}
\end{align}
with $S_{ij} = \Ket{i}\Bra{j}~; i,j \in \{ g,a,s,e \}$; $\Ket{g}$, $\Ket{a}$, $\Ket{s}$ and $\Ket{e}$ being the ground, antisymmetric, symmetric and doubly excited state of our joint system, respectively. 

\section{SE rate and cross Lamb shift terms for dipole perpendicular to the surface}

In the case of a dipole perpendicular to the surface, expressions for the cross Lamb shift term and SE rate similar to the ones used in section \ref{image:dot:approach} can be derived from first principles as well, arriving at the expressions
\begin{align}\label{Fimage}
\begin{split}
\mathcal{F}_{12}(q \Delta r) = 3\left( - \frac{\cos(q \Delta r)}{(q \Delta r)^2} + \frac{\sin(q \Delta r)}{(q \Delta r)^3} \right)~,
\end{split}
\end{align}
and
\begin{align}
\begin{split}
\mathcal{G}_{12}(q \Delta r) = -3\left(\frac{\sin(q \Delta r)}{(q \Delta r)^2} + \frac{\cos(q \Delta r)}{(q \Delta r)^3}\right)~, 
\end{split}
\end{align}
instead of the ones used in section \ref{image:dot:approach}.

\clearpage
\bibliography{references}
\bibliographystyle{unsrt}

\end{document}